\def\be{\begin{equation}}
\def\ee{\end{equation}}
\def\bea{\begin{eqnarray}}
\def\eea{\end{eqnarray}}
\def\b{\bibitem}
\documentclass{ws-p9-75x6-50}

\begin{document}
\title{Superconductivity and Quantum Phase Transitions in Weak
       Itinerant Ferromagnets}

\author{T.R. Kirkpatrick}
\address{Insitute for Physical Science and Technology, and Department of 
         Physics, University of Maryland, College Park, MD 20742}

\author{Thomas Vojta}
\address{Department of Physics, University of Oxford, 1 Keble Rd, Oxford OX1
         3NP, UK, and \\
         Institut f{\"u}r Physik, TU Chemnitz, D-09107 Chemnitz, FRG}

\author{D. Belitz}
\address{Department of Physics and Materials Science Institute, University 
         of Oregon, Eugene, OR 97403, USA}

\author{R. Narayanan}
\address{Department of Physics, University of Oxford, 1 Keble Rd, Oxford OX1
         3NP, UK}
\maketitle
\abstracts{It is argued that the phase transition in low-$T_c$ clean
itinerant ferromagnets is generically of first order, due to correlation
effects that lead to a nonanalytic term in the free energy. A tricritical
point separates the line of first order transitions from Heisenberg critical
behavior at higher temperatures. Sufficiently strong quenched disorder
suppresses the first order transition via the appearance of a critical
endpoint. A semi-quantitative discussion is given in terms of recent
experiments on MnSi and UGe$_2$.
It is then shown that the critical temperature for spin-triplet, p-wave
superconductivity mediated by spin fluctuations is generically much higher
in a Heisenberg ferromagnetic phase than in a paramagnetic one, due to the
coupling of magnons to the longitudinal magnetic susceptibility.
This qualitatively explains the phase diagram recently observed in UGe$_2$
and ZrZn$_2$.}

\section{Introduction}
\label{sec:1}

In this paper we convey two messages: In the first part we argue that 
in sufficiently clean samples, and at sufficiently low temperatures,
the ferromagnetic phase transition in itinerant electron sytems is
generically of first order. In the second part, we provide a physical
explanation for the observed structure of the phase diagram in UGe$_2$
and ZrZn$_2$, where superconductivity is observed to coexist with
ferromagnetism.

\subsection{Multicritical points in Itinerant Ferromagnets}
\label{subsec:1.1}

The thermal paramagnet-to-ferromagnet transition at the Curie temperature
is usually regarded as a prime example of a continuous or second order phase
transition: Upon cooling, the magnetization increases continuously from zero
above the Curie temperature, to finite values below the Curie point. For
materials with high Curie temperatures this behavior is well established.

\begin{figure}[t]
\epsfxsize=13pc 
\centerline{\epsfbox{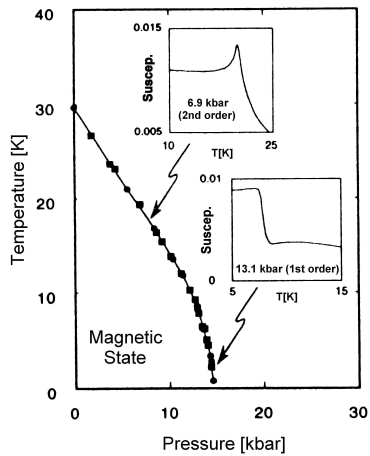}} 
\caption{Phase diagram of MnSi. The insets show the behavior of
   the susceptibility close to the transition.
  (After Ref. \protect\citelow{Pfleiderer_et_al}). }
\label{fig:1}
\end{figure}

Recently there has been a considerable interest in the corresponding quantum
phase transition of itinerant electrons, that takes place at zero
temperature as a function of some non-thermal control parameter. Understanding
the quantum phase transition is important, since it controls large parts of
the critical behavior that is observable in systems with nonzero, but low,
Curie temperatures. There is experimental evidence that 
in sufficiently clean itinerant ferromagnets the phase transition is 
discontinuous, or of first
order, provided that the Curie temperature is low enough. Specific systems
exhibiting this behavior include, MnSi\cite{Pfleiderer_et_al} and 
UGe$_2$.\cite{Saxena_et_al} In both of
these systems the transition temperature can be tuned to zero by varying the
pressure. It is found that there is a critical pressure, $p_{c}$, above
which the ferromagnetic phase transition is discontinuous. An example of a 
phase diagram is shown in Fig.\ \ref{fig:1}.

In the first part of this paper, Sec.\ \ref{sec:2},
we review a general reason for why we expect
all sufficiently clean itinerant electron systems to have a discontinuous
ferromagnetic transition, if the transition temperature is low enough.

\subsection{Coexistence of Ferromagnetism and Superconductivity}
\label{subsec:1.2}

At first glance, and according to conventional wisdom, ferromagnetism 
and superconductivity seem incompatible with
one another. For superconductivity with conventional s-wave pairing,
the large internal magnetic field inside a magnet would make this
singlet pairing energetically very costly. Triplet p-wave pairing, with the
spins aligned with the magnetism, is a possibility, but since
superconductors tend to expel magnetic flux, one is, again, naively led to the
conclusion that superconductivity and ferromagnetism are likely
incompatible.

Nevertheless, recent experiments indicate that in some very pure systems,
and at very low temperatures,
ferromagnetism and superconductivity can coexist, with the
same electrons that cause the magnetism also responsible for the
superconductivity. So far this phenomenon has been observed in two systems,
UGe$_2$\cite{Saxena_et_al} and ZrZn$_2$,\cite{Pfleiderer_et_al_2} 
and it is believed to be generic.

\begin{figure}[t]
\epsfxsize=18pc 
\centerline{\epsfbox{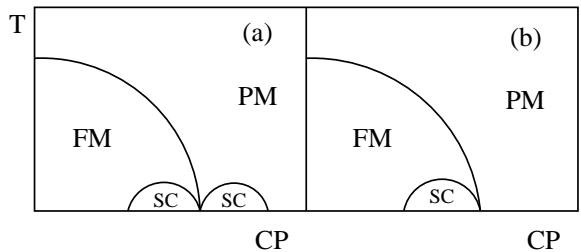}} 
\caption{Schematic phase diagram showing the paramagnetic (PM), ferromagnetic
 (FM), and superconducting phases (SC) in a temperature (T) - control parameter
 (CP) plane. (a) shows the qualitative prediction of paramagnon
 theory, Ref.\ \protect\citelow{FayAppel}, and 
 (b) qualitatively shows the phase diagram
 as observed in UGe$_2$, Ref.\ \protect\citelow{Saxena_et_al}.}
\label{fig:2}
\end{figure}

These experiments raise a number of obvious questions. First, what is the
nature of the superconducting pairing? Does it have s-wave, p-wave, or some other
symmetry? What is the nature of the superconducting state? The Meissner effect
leads one to believe that the superconducting state must be inhomogeneous.
On a more microscopic level, what is the pairing mechanism? In analogy with
the phonon mechanism for conventional superconductivity, it was argued
theoretically already in the 1960's that magnetic fluctuations could induce
pairing.\cite{AndersonBrinkman} 
These theories led to phase diagrams with
superconducting phases that appeared
more or less symmetric around the ferromagnetic phase boundary.\cite{FayAppel}
The basic idea behind these theories was 
that the magnetic fluctuations are largest near a continuous magnetic phase 
transition, and if a fluctuation induced superconducting state is to be
obtained, then it will most likely exist near the magnetic phase boundary.
These old theories are in conflict both with our suggestion that the
low-temperature
ferromagnetic transition is discontinuous in the very 
pure systems needed to
observe superconductivity, and with the experimental observation that the
superconducting state is observed only on the ferromagnetic side of the
magnetic phase boundary. A schematic phase diagram is shown in 
Fig.\ \ref{fig:2}.

In the second part of this paper, Sec.\ \ref{sec:3}, we review a general 
pairing mechanism that
leads to the conclusion that one should expect a p-wave paired
superconducting state to effectively exist only on the ferromagnetic side of
the phase boundary, consistent with the experimental observations.

\section{Ferromagnetism in clean itinerant systems}
\label{sec:2}

On general grounds, Landau\cite{LL} 
argued that as a function of the magnetization
$m$, the free energy for small $m$ is of the form 
\be
F = t\,m^2 + u\,m^4 + O(m^6)\quad.
\label{eq:1}
\ee
Within Landau theory, this holds independently of whether one deals with a
magnet at zero or finite temperature. In Eq.\ (\ref{eq:1}),
$t$ is some dimensionless distance from the critical point, and $u$ is
assumed to be a positive constant. This equation implies a continuous
paramagnetic-to-ferromagnetic phase transition at $t=0$ ($t<0$ describes the
ferromagnetic phase and $t>0$ the paramagnetic one) with mean-field
critical exponents. It is well
known that in general fluctuations effects modify this result for systems
whose spatial dimensionality is less than an
critical upper dimension, $d_{c}^{+}$, resulting in a continuous phase
transition with critical exponents that depend on the spatial
dimension as well as the dimension of the order parameter.\cite{Fisher} 
In effect,
Landau's reasoning breaks down because the free energy is not an analytic
function of the magnetization at the critical point. Another well-known, but
non-generic mechanism that can invalidate Eq.\ (\ref{eq:1}) is that for
some systems the coefficient $u$ in Eq.\ (\ref{eq:1}) can be negative.
In that case one needs to keep the term of $O(m^6)$ in the free energy,
and the transition is discontinuous, or of first order. The point in the
phase diagram where $u$ changes sign, as a
function of some microscopic parameter, is
known as a tricritical point.

This general picture was not expected to be modified if the
transition took place at zero rather than finite temperatures. Indeed, the
only expected change was that the value $d_{c}^{+}$ would be changed for a
zero-temperature transition compared to its thermal counterpart.\cite{Hertz}
This expectation turned out to be incorrect, at least for itinerant 
ferromagnets. The basic reason for its breakdown
is that, in a zero temperature itinerant electron system, soft modes
that are unrelated to the critical order parameter (OP) or magnetization
fluctuations couple to the latter. This leads to an effective long-range
interaction between the OP fluctuations, which in turn leads to a
nonanalytic magnetization dependence of the free energy, unrelated to the
nonanlyticities due to critical fluctuations.\cite{us_fm_dirty} 
In disordered systems, the
additional soft modes are the same `diffusons' that cause the so-called weak
localization effects in paramagnetic metals.\cite{LeeRama}
In clean systems, they are ballistic modes that lead to
corrections to Fermi liquid theory.\cite{us_fm_clean}

To see these effects we consider the functional form of the free energy of a
bulk itinerant ferromagnet at finite temperatures, in the absence of
quenched disorder (for a discussion including the effects of disorder, see
Ref.\ \citelow{us_fm_dirty}). In Ref.\ \citelow{us_fm_clean} 
it was shown that at $T=0$ there is a contribution to the
free energy from the additional soft modes that is 
schematically given by an integral over a
frequency $\omega $ and wavenumber $k$, 
\be
f(m) = -m^4\int_0^{\Lambda}dk\,k^{d-1}\int_0^{\infty}d\omega\,
       \frac{1}{[(\omega + k)^2 + m^2]^{2}}\quad.
\label{eq:2}
\ee
Here $\Lambda $ is a cutoff, and the crucial sign of this contribution will be
discussed below. Equation\ (\ref{eq:2}) gives 
$f(m\rightarrow 0)\propto -m^{d+1}$ for $1<d<3$, and 
$f(m\rightarrow 0)\propto m^4\ln m$ in $d=3$. From now on, we
restrict ourselves to the $d=3$ case. The leading effect of a nonzero
temperature is adequately represented by replacing $\omega \rightarrow
\omega +T$ in the integrand. The net result is that at low temperatures the
Landau free energy given by Eq.\ (\ref{eq:1}) is generalized 
to\cite{us_1st_order}
\be
F = t\,m^2 + v\,m^4\ln (m^2 + T^2) + u\,m^4 + O(m^6)\quad. 
\label{eq:3}
\ee
The sign of the coefficient $v$ in Eq.\ (\ref{eq:3}) 
warrants some attention. The
derivation of the term to leading order in the electron-electron interaction 
($O(\Gamma_t^2)$, with $\Gamma_t$ a spin-triplet interaction amplitude) 
yields $v>0$.\cite{us_1st_order} We further note
that $v>0$ indicates a decrease in the tendency towards magnetism due to
correlation effects. This can be seen by remembering that $F$ can be related
to the magnetic susceptibility. It is well known that correlation effects in
general decrease the tendency toward ferromagnetism, so our perturbative
results seems to be the generic one. In what follows we therefore assume 
that $v>0$.

We next analyze the equation of state that follows from Eq.\ (\ref{eq:3}).
At $T=0$, the transition is clearly of first order, since $m^{4}\ln m<0$ for
small $m$. \ The transition occurs at $t=v\exp [-(1+u/v)]>0$,
and the magnetization at the
transition is $m=\exp [-(1+u/v)/2]$. Since we have truncated the OP
expansion in Eq.\ (\ref{eq:3}), these results are exact only for 
$u/v>>1$, but similar
results are expected when this inequality is not satisfied.

At nonzero $T$, the free energy is an analytic function of $m$, but for
small $T$ the coefficients in an $m$-expansion become large. There is a
tricritical point at $T_{tc}=\exp (-u/2v)$, with a first order transition
for $T<T_{tc}$, and a line of Heisenberg critical points for $T>T_{tc}$.
These results, at $T=0$ and $T>0$, lead to a phase diagram as given in
Fig.\ \ref{fig:1}. We stress that, since the nonanalytic term in 
Eq.\ (\ref{eq:3}) is due to the
long-wavelength excitations in the itinerant electron system, the phase
diagram is expected to be generic.

If one adds a finite amount of quenched disorder, the predicted phase 
diagram becomes quite complicated. Most importantly, a sufficient amount 
of quenched disorder
causes the transition to become continuous. This behavior and the associated
critical endpoint (and multicritical points) as well as other features of
the phase diagram are discussed in Ref.\ \citelow{us_1st_order}.

\section{Spin-fluctuation induced superconductivity in ferromagnets}
\label{sec:3}

In order to study ferromagnetic spin-fluctuation induced superconductivity,
we choose an OP field for the superconductivity as
${\cal F}(x,y) = \psi_{\uparrow}(x)\psi_{\uparrow}(y)$, 
assuming that the magnetization
in the ferromagnetically ordered phase is in the z-direction. Here 
$\psi_{\sigma}(x)$ is an electronic field with spin index $\sigma$ and 
space-time index $x$. The OP, i.e. the expectation value 
$\langle{\cal F}(x,y)\rangle \equiv F(x-y)$, 
is the anomalous Green's function. At this point the
orbital symmetry of the OP is unspecified, but we eventually choose p-wave
pairing. Note, that in choosing the above OP we have assumed
a particular form of triplet pairing
which, as noted in the Introduction, 
is the most likely superconducting state. For simplicity, and to get a first
handle on the theory, we will
also assume that the superconducting state is homogeneous. This assumption, as 
was also noted in the Introduction, is questionable, and its consequences
warrant further investigation. Finally, we will not be concerned that
ordinary classical critical phenomena might lead to nontrivial magnetic
fluctuations close to the magnetic phase boundary because, as pointed out in
Sec.\ \ref{sec:2} above, we expect a discontinuous magnetic phase
transition at low temperatures. Further, as we will see below, even if the
predicted tricritical point was at inaccessibly low temperatures, so that
the transition were effectively always continuous, large scattering very
near the transition suppresses the superconducting state very close to the
continuous phase boundary anyway.\cite{FayAppel,LevinValls,RoussevMillis}

Using a field theoretic approach, and working to leading order in magnetic
fluctuations, we have derived coupled equations of motion for $F$
and the normal Green function, 
$G_{\sigma}(x-y) = \langle{\bar\psi}_{\sigma}(x)\,\psi_{\sigma}(y)\rangle$, 
that lead to an equation of state
for the superconducting OP. 
In an approximation analogous to 
Eliashberg theory for conventional superconductivity, we obtain for the
linearized gap equation that determines the 
superconducting critical temperature $T_c$,
\bea
\Delta(p)&=&\Gamma_{t}T\sum_{k}\chi_{L}(p-k)\vert G_{\uparrow }(k)\vert^2
            \Delta (k)\quad,
\label{eq:4}\\
G_{\sigma}(p)&=&1/\left[i\omega_n - \xi_{\bf p} - \Sigma_{\sigma}(p)\right]
               \quad,
\label{eq:5}\\
\Sigma_{\sigma}(p)&=&\Gamma_t T\sum_k\left[\chi_L(p-k)\,G_{\sigma}(k) 
            + 2\chi_T(p-k)\,G_{-\sigma}(k)\right]\quad.
\label{eq:6}
\eea
Here we work in Fourier space, with $p = ({\bf p},i\omega_n)$
comprising the momentum and the Matsubara frequency, and
$\sigma = +,- \equiv \uparrow,\downarrow$.
$\xi_{\bf p} = \epsilon_{\bf p} - \mu$ is the bare
quasiparticle spectrum minus the chemical potential, $\Sigma$ is the normal
self-energy, $\Gamma_t$ is the spin-triplet interaction amplitude, 
$\chi_{L,T}$ are the longitudinal and transverse magnetic susceptibilities,
respectively, and $\Delta$ is the anomalous self-energy. Note that in the
paramagnetic phase, $\chi_L = \chi_T$ and $G_{\uparrow} = G_{\downarrow}$.

We have solved Eqs.\ (\ref{eq:4}) - (\ref{eq:6}) in a simple
McMillan-type approximation. We find for the superconducting transition
temperature\cite{us_UGe_2}
\be
T_c = T_0(t)\,\exp\left[-(1 + d_0^L + 2d_0^T)/d_1^L\right]\quad.
\label{eq:6a}
\ee
Here $T_0(t)$ is some measure of the
magnetic excitation energy. Following Ref.\ \citelow{FayAppel}, we use the
prefactor of $\vert t\vert$ in Eqs.\ (\ref{eq:7}) and (\ref{eq:8}) below,
\be
T_0(t) = T_0\,[\Theta(t)\,t + \Theta(-t)\,5\vert t\vert/4]\quad,
\label{eq:6a'}
\ee
with $T_0$ a microscopic temperature scale that is related to the Fermi
temperature (for free electrons) or a band width (for band electrons).
This qualitatively reflects the suppression of the
superconducting $T_c$ near the FM transition due to effective
mass effects\cite{LevinValls,FayAppel,RoussevMillis}.

Specializing to the p-wave case, the $d_{0,1}^{L,T}$ read
\bea
d_1^L&=&\left(\Gamma_t N_F^{\uparrow}/(k_F^{\uparrow})^2\right)
   \int_{0}^{2k_F^{\uparrow}} dk\,k\,\left(1 - \left(k^2/2(k_F^{\uparrow})^2
      \right)\right)\,D_L(k,i0)\ ,
\label{eq:6b} \\
d_0^L&=&\left(\Gamma_t N_F^{\uparrow}/(k_F^{\uparrow})^2\right)
   \int_{0}^{2k_F^{\uparrow}} dk\,k\,D_L(k,i0)\quad,
\label{eq:6c}\\
d_0^T&=&\left(\Gamma_t N_F^{\uparrow}/(k_F^{\uparrow})^2\right)
   \int_{k_F^{\uparrow}-k_F^{\downarrow}}^{k_F^{\uparrow}+k_F^{\downarrow}}
      dk\,k\,D_T(k,i0)\quad.
\label{eq:6d}
\eea
$k_F^{\uparrow} (k_F^{\downarrow})$ are the Fermi wavenumbers for the
up (down)-spin Fermi surface, and $N_F^{\uparrow}$ is the density of states
at the up-spin Fermi surface. In the paramagnetic phase,
$k_F^{\uparrow} = k_F^{\downarrow} \equiv k_F$. $D_{L,T}(q)$ are the
longitudinal and transverse (para)magnon propagators, which are related to
the electronic spin susceptibility $\chi$ via
$D_{L,T}(q) = \chi_{L,T}(q)/2N_F$, with $N_F$ the density of states at
the Fermi level in the paramagnetic phase. In order to perform the integrals
we need to specify the susceptibilities. We use the
expressions that were derived in Ref.\ \citelow{us_fm_mit}, with one
crucial modification that we will discuss below.
In the paramagnetic phase, in the limit of small wavenumbers, 
\be
D_{L,T}({\bf q},i0) = 1/\left[t + b_{L,T}({\bf q}/2k_F)^{2}\right]\quad, 
\label{eq:7}
\ee
with $b_{L}=b_{T}$ and $k_{F}$ the Fermi wavenumber. In the Gaussian
approximation of Ref.\ \citelow{us_fm_mit}, 
$b_{L}=b_{T}=1/3.$ However, more generally we note
that Eq.\ (\ref{eq:7}) 
is expected to be a generic form in the long wavelength limit
with the $b^{\prime }s$ of $O(1)$. Similarly, in the long wavelength limit,
in the ferromagnetic phase, 
\bea
D_L({\bf q},i0)&=&1/\left[5\vert t\vert/4 + b_L({\bf q}/2k_F)^2\right]\quad, 
\label{eq:8}\\
D_T(q)&=&\frac{\Delta/4\epsilon_F}{(1-t)^2}\left(\frac{1}{i\Omega
        +(\Delta/2\epsilon_F)\,b_T({\bf q}/2k_F)^2} + {\rm c.c.}\right)\quad,
\label{eq:9}
\eea
with $\Delta $ the band splitting energy. The factor $5/4$ in 
Eq.\ (\ref{eq:8}) can be
traced back to the fact that the particle number is typically held fixed in
experiments. For $0<\Delta <n\Gamma _{t}$, $\Delta $ is related to the
magnetization by $m=\mu _{B}\Delta /\Gamma _{t}$, with $n$ the electron
density.

Let us discuss the propagators in the magnetic state. The
form of the transverse propagator, Eq.\ (\ref{eq:9}), is known 
to be asymptotically exact in
the long-wavelength limit, where the spectrum describes the spin-wave or
magnon excitations. Equation (\ref{eq:8}) for the longitudinal propagator,
on the other hand, is an RPA or Landau-type approximation that was used
in previous theories of magnetic fluctuation induced 
superconductivity.\cite{FayAppel} It is easy to see that this approximation
leads to a
superconducting phase diagram that is more or less symmetric with respect 
to the magnetic phase boundary. First, the longitudinal susceptibilities are
roughly the same one either side of the transition. Second, the transverse
propagators, which are fundamentally different in the two phases, only
weakly couple to the superconducting gap equation well inside the magnetic
phase where $m$ is not too small. The conclusion is that apart from simple
factors, the linearized gap equation is basically the same on both sides of
the magnetic phase transition.

\begin{figure}[t]
\epsfxsize=10pc 
\centerline{\epsfbox{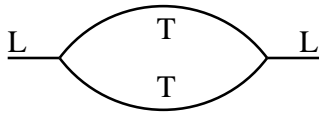}} 
\caption{Mode-mode coupling contribution to the longitudinal (L) propagator
 $D_L$ from the transverse (T) ones.}
\label{fig:3}
\end{figure}

We next consider the longitudinal propagator in
the ferromagnetic phase in more detail. In a Heisenberg ferromagnet (or in
any magnet with a continuous rotation symmetry in spin space), the
transverse spin waves or magnons are massless and couple to the longitudinal
susceptibility $\chi _{L}$.\cite{BrezinWallace}
This effect is most easily illustrated within a
nonlinear sigma-model description of the ferromagnet,\cite{ZJ}
which treats the order
parameter ${\bf m}$ as a vector of fixed length $m$, and parametrizes it
as ${\bf m} = m(\pi_1(x),\pi_2(x),\sigma(x))$, with 
$\sigma^2 + \pi_1^2 + \pi_2^2 = 1$, and $m$ the magnetization.
The diagonal part of the transverse or $\pi$ propagator, 
$\langle\pi_i\pi_i\rangle = (m^2/2N_F)D_T$, is proportional to the transverse
propagator $D_{T}$, and the off-diagonal part has been 
calculated in Ref.\ \citelow{BKMV}.
The longitudinal propagator, 
$D_L = (m^2/2N_F)\left\langle\sigma(x)\sigma(y)\right\rangle$,
can be expanded in a series of $\pi$-correlation functions as,
\be
\left\langle\sigma(x)\sigma(y)\right\rangle = 1 
   - \left\langle\pi_i(x)\pi_i(x)\right\rangle
   + \frac{1}{4}\left\langle\pi_i(x)\pi_i(x)\pi_j(y)\pi_j(y)\right\rangle
   + \ldots
\label{eq:10}
\ee
where the repeated indices are summed over. At one-loop order, the term of
order $\pi ^{4}$ yields the diagram shown in Fig.\ \ref{fig:3}. 
Power counting shows
that at nonzero temperature, and for dimensions $d<4$, this contribution
causes the homogeneous longitudinal susceptibility to diverge everywhere in
the ferromagnetic phase, so $\chi_L$ is fundamentally different in
the ferromagnetic phase than in the paramagnetic one.\cite{BrezinWallace}
Ultimately this
implies the superconducting transition temperature can be very different in
the two phases, and it is this aspect that previous theories missed.

More generally, this one-loop contribution, together with the zero-loop one,
Eq.\ (\ref{eq:8}), 
yields a functional form for $D_L$ in the ferromagnetic phase that
is asymptotically exact at small wavenumbers. This diagram has no analog in
the paramagnetic phase, while all other renormalizations of the propagators
will give comparable contributions in the ferromagnetic and paramagnetic
phases. It is therefore reasonable to calculate $T_c$ based one this
one-loop result in the ferromagnetic phase, and compare it to the zero-loop
calculation in the paramagnetic phase.

\begin{figure}[t]
\epsfxsize=14pc 
\centerline{\epsfbox{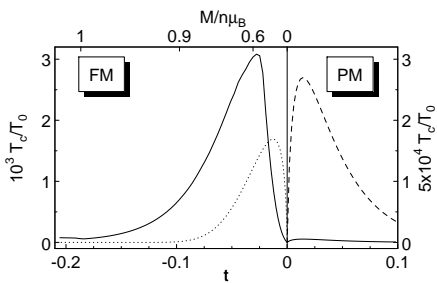}} 
\caption{Superconducting $T_c$ (solid curve, left scale) as a function
 of the distance from the critical point $t$, and the magnetization $M$.
 The dashed line (right scale) shows $T_c$ in the paramagnetic phase scaled by a
 factor of 50, and the dotted curve (right scale) is the result in the 
 ferromagnetic phase without the mode-mode coupling effect. 
 From Ref.\ \citelow{us_UGe_2}.}
\label{fig:4}
\end{figure}

\begin{figure}[t]
\epsfxsize=14pc 
\centerline{\epsfbox{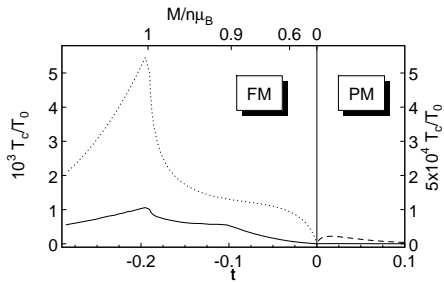}} 
\caption{Same as Fig.\ \ref{fig:2}, but for different parameter values
         (see the text). From Ref.\ \citelow{us_UGe_2}.}
\label{fig:5}
\end{figure}

In the McMillan approximation noted above, two examples of the resulting
phase boundaries for superconductivity in the paramagnetic and ferromagnetic
phases are shown in Figs.\ \ref{fig:4} and \ref{fig:5}. 
In both figures, the characteristic
temperature $T_0$ is given by either the Fermi temperature or a band
width, depending on the model considered. The magnetization $m$ has been
scaled with $\mu_{B}n$, with $\mu_{B}$ the Bohr magneton. The solid line
represents the superconductivity $T_c$ in the ferromagnetic phase as a
function of the distance $t$ from an assumed continuous ferromagnetic
critical point. Since the transition is discontinuous, the region very close
to the point $t=0$ should be ignored. The dashed line shows the result in the
paramagnetic phase scaled by a factor of 50 (right hand scale), and the
dotted curve in the ferromagnetic phase (also scaled by a factor of 50,
right-hand scale) represents the result that is obtained in the
ferromagnetic phase upon neglecting the mode coupling contribution to 
$\chi_L$ given by Fig.\ \ref{fig:3}. In Fig.\ \ref{fig:4} the values 
$b_L = 0.23$ and $b_T = 0.4$
are used, while in Fig.\ \ref{fig:5} 
the values $b_L = b_T = 1$ are used. In both
cases, note that the maximum $T_c$ in the ferromagnetic phase is $50$
to $100$ times higher than in the paramagnetic phase.

We conclude that for reasonable parameter values, theoretically the
effective superconducting phase diagram is given by Fig.\ \ref{fig:2}, 
consistent with current experimental observations.

\section{Discussion}
\label{sec:4}

We conclude with a summary of our results, and then briefly
discuss several open questions.

We have made two distinct general points. The first result was that
clean itinerant electronic systems will in general
have a tricritical point for the ferromagnetic phase transition at low
temperatures. As a corollary, the zero temperature ferromagnetic
transition in clean itinerant systems is generically of first order. 
This result continues to hold for weakly disordered
systems, but no quantitative results are available for the amount of
disorder that will destroy the first order phase transition.
Once the tricritical point has been destroyed by the disorder, the
critical exponents at the second order phase transition at finite 
temperatures are the known classical Heisenberg exponents,\cite{ZJ}
and at zero temperature, they have recently
been exactly determined in Ref.\ \citelow{us_fm_dirty}. 
The second result was that longitudinal
fluctuations are intrinsically larger in the ferromagnetic phase than in the
paramagnetic phase. This is a crucial point for magnetic fluctuation induced
superconductivity. Simple estimates show that the critical temperature for
this type of superconductivity in the ferromagnetic phase 
can easily be fifty times larger than in the paramagnetic one.
All of these results are consistent with current experimental
observations.

The most intriguing open questions concern the nature of the magnetic
fluctuation induced superconducting state, and an understanding of the phase
diagram for all temperatures and magnetic fields. As already noted, one
expects an inhomogeneous superconducting state. This point has not
yet been discussed theoretically for pairing mechanisms that are both
electronic in origin, and sensitive to internal magnetic field effects.
Another interesting question is whether there are numerous superconducting
phases as a function of temperature (and external magnetic
fields). Again, since the pairing mechanism is expected to be electronic in
origin, and itself sensitive to superconductivity, it is easy to imagine
additional superconducting states appearing inside the superconducting
phase, as the temperature is lowered. Similarly, the concept of transverse
and longitudinal critical external magnetic fields needs to be worked out
for these superconducting states.

\section*{Acknowledgments}
We would like to acknowledge support by the National Science Foundation
through grant Nos. DMR-98-70597 and DMR-99-75259, by the DFG, grant No.
Vo659/3, and by the EPSRC, grant No. GR/M 04426.

\end{document}